\documentclass[prb,twocolumn,showpacs,amsmath,amssymb,superscriptaddress]{revtex4}
\usepackage{graphicx}
\usepackage{dcolumn}
\usepackage{bm}
\usepackage{epsfig}
 
\begin{document}

\title{Superconductivity and fluctuating magnetism in quasi two-dimensional 
$\kappa$-(BEDT-TTF)$_{2}$Cu[N(CN)$_{2}$]Br probed with implanted muons}

\author{T. Lancaster}
\email{t.lancaster1@physics.ox.ac.uk} 
\affiliation{Oxford University Department of Physics, Clarendon Laboratory,  
Parks Road, Oxford, OX1 3PU, UK
}
\author{S.J. Blundell} 
\affiliation{Oxford University Department of Physics, Clarendon Laboratory,  
Parks Road, Oxford, OX1 3PU, UK
}

\author{F.L. Pratt}
\affiliation{ISIS Facility, Rutherford Appleton Laboratory, Chilton, Oxfordshire OX11 0QX, UK}

\author{J.A. Schlueter}
\affiliation{Materials Science Division,
Argonne National Laboratory, 9700 South Cass Ave., Argonne, IL 60439, US}

\date{\today}

\begin{abstract}
A muon-spin relaxation ($\mu^{+}$SR) investigation is presented for the 
molecular superconductor $\kappa$-(BEDT-TTF)$_{2}$Cu[N(CN)$_{2}$Br]. 
Evidence is found for low-temperature phase-separation, with only a fraction of the sample showing a superconducting signal,
even for slow cooling. Rapid cooling reduces the superconducting fraction still further. 
For the superconducting phase, the in-plane penetration depth is measured to be 
$\lambda_{\parallel} =  0.47(1)~\mu$m and evidence is seen for a vortex decoupling transition in applied fields above 40~mT. 
The magnetic fluctuations in the normal state produce 
Korringa behavior of the muon spin relaxation rate below 100~K, a precipitous drop in relaxation rate 
is seen at higher temperatures and an enhanced local spin susceptibility occurs just above $T_{c}$.
\end{abstract}
\maketitle

\section{Introduction}

The quasi two-dimensional molecular series $\kappa$-(BEDT-TTF)$_{2}X$ 
exemplifies the complex interplay of collective phenomena that occur in 
correlated electron systems\cite{ishiguro,powell}. In particular, the proximity of 
superconducting $\kappa$-(BEDT-TTF)$_{2}$Cu[N(CN)$_{2}$]Br to a Mott transition, along with 
possible pseudogap physics and  molecular disorder effects, has led to this 
material 
being extensively studied in recent years\cite{sano,taylor,yusuf,nam,wolter}. Despite this
intense interest, several
experimental observations in this system lack a conclusive explanation
\cite{powell}
and more experimental study is warranted.
In this paper we present a muon-spin relaxation ($\mu^{+}$SR) investigation of 
$\kappa$-(BEDT-TTF)$_{2}$Cu[N(CN)$_{2}$]Br. Our focus is the penetration depth in 
the vortex state for $T<T_{c}$ and the fluctuations of the local magnetic field 
in the normal state. 

\begin{figure}
\begin{center}
\epsfig{file=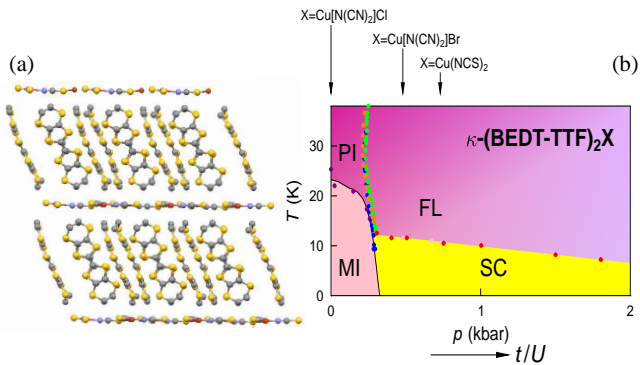,width=\columnwidth}
\caption{(a) Layered 
structure of the $\kappa$-(BEDT-TTF)$_{2}X$ system. 
(b) Phase diagram of the system showing the proximity of the $X=$Cu[N(CN)$_{2}$]Br
salt to the Mott insulator phase. (MI=Mott Insulator, PI=Paramagnetic insulator, 
SC=Superconductor, FL=Fermi liquid.) (After Ref.~\onlinecite{nam}).
\label{picture}}
\end{center}
\end{figure}

The layered structure of the $\kappa$-(BEDT-TTF)$_{2}X$ family of
organic molecular metals is shown in Fig.~\ref{picture}(a). 
The BEDT-TTF molecules dimerize, forming molecular
units which stack on a triangular lattice in two-dimensional planes. 
The removal of one electron from each BEDT-TTF dimer causes the tight-binding band to be half filled
and in order
to balance the charge
layers of anion, X, are located between the partially oxidized
BEDT-TTF sheets.
The properties of the system are strongly dependent on the transfer integral $t$ 
which is controlled
by the dimer separation and 
may be manipulated by applying pressure or by changing the identity of the
anion\cite{powell}.  
The phase diagram of the family
[Fig.~\ref{picture}(b)] shows that the system may be tuned from
Mott insulator through superconductivity into a normal metallic state 
as a function of $t/U$, where $U$
is the on-site Coulomb repulsion (which is a property of the dimer and is almost independent of
the anion or pressure). 
Importantly, this tuning is achieved without the need for
chemical doping (in contrast to the cuprates) 
and therefore minimizes structural disorder effects.
At small $t/U$ Coulomb correlations dominate and $\kappa$-(BEDT-TTF)$_{2}X$
with $X=$Cu[N(CN)$_{2}$]Cl is a Mott insulator \cite{kagawa}. As pressure is increased or $X$ is changed to 
$X=$Cu[N(CN)$_{2}$]Br there is an insulator-to-superconductor transition \cite{williams}. 
The $X=$Cu(NCS)$_{2}$ compound exhibits a larger $t/U$ still and remains a superconductor 
with a slightly depressed $T_{c}$. 

\begin{figure}
\begin{center}
\epsfig{file=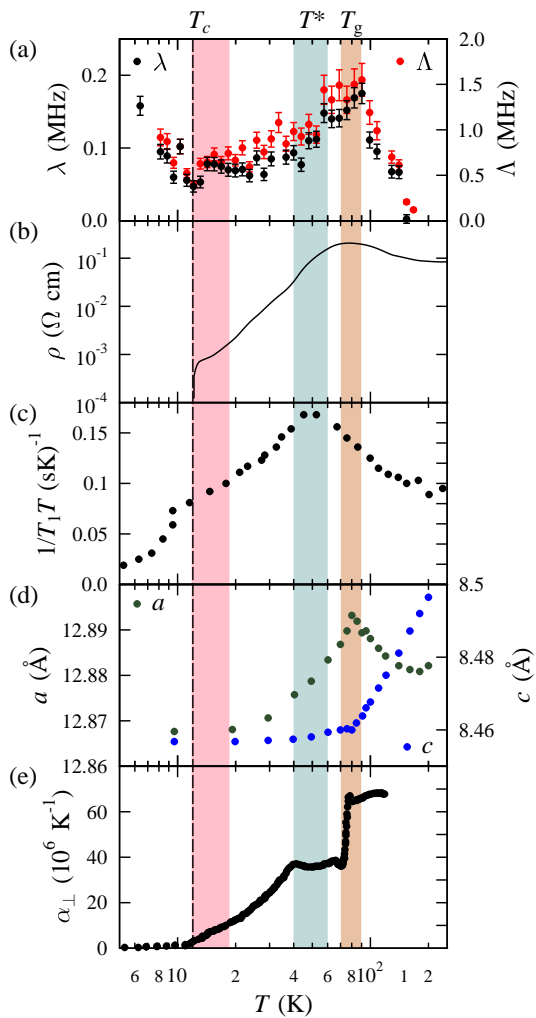,width=7cm}
\caption{Comparison of several experimental properties with regions~I ($T\approx 80$~K), II 
($T \approx 50~K$) and III ($T_{c}< T \lesssim 18$~K) shaded. (a) LF $\mu^{+}$SR relaxation rates (this work); (b) Resistivity \cite{yu};
(c) NMR $1/T_{1}T$ \cite{miyagawa}; (d) lattice constants \cite{wolter}; (e) thermal expansion
coefficient perpendicular to the layers\cite{muller}.
\label{master}}
\end{center}
\end{figure}

Superconducting $\kappa$-(BEDT-TTF)$_{2}$Cu[N(CN)$_{2}$]Br ($T_{c}\approx 12$~K)
displays several properties [Fig.~\ref{master}] whose explanation 
remains obscure. These may be broadly grouped three temperature regions. 
In region~I ($T\approx 80$~K) there are sharp changes in 
the temperature dependence of the lattice constants\cite{wolter} [Fig.~\ref{master}(d)] 
and, above 70~K, a rapid increase in the thermal expansion coefficient perpendicular to the planes\cite{muller} 
$\alpha_{\perp}$ [Fig.~\ref{master}(e)]. These 
coincide with a rounded maximum in the resistivity\cite{yu} $\rho$ [Fig.~\ref{master}(b)]. The
properties of this region have been linked to a glass-like freezing of terminal ethylene
groups on the BEDT-TTF molecules\cite{muller} around $T_{\mathrm{g}}=77$~K, 
although is has been suggested \cite{wolter} that 
this freezing alone cannot account for the observed structural changes. 
In region~II ($T \approx 50$~K) NMR measurements \cite{miyagawa,powell}
yield a maximum in $1/T_{1}T$  around $T^{*}\approx 50$~K [Fig.~\ref{master}(c)].
Upon cooling through this region there is a crossover in transport
properties from ``bad metal'' behavior to
a more conventional Fermi-liquid regime\cite{powell} and  a plateau
in $\alpha_{\perp}$ is also seen [Fig.~\ref{master}(e)]. These phenomena
have been linked to pseudogap physics although the details remain unclear\cite{powell}. 
In region~III ($T_{c}< T \lesssim 18$~K) a  vortex Nernst signal is observed \cite{nam}.
It was argued\cite{nam} that the proximity of 
the $X=$Cu[N(CN)$_{2}$]Br material to the Mott state in the phase diagram 
in Fig.~\ref{picture}(b), results in vortex fluctuations persisting
above $T_{c}$ and, furthermore, this remnant of fluctuating 
superconductivity may be consistent with the occurrence of phase fluctuations 
in the superconducting order parameter close to the Mott boundary. 

This paper is structured as follows: after discussing experimental details in section~II we
present the results of transverse field muon-spin relaxation 
measurements of the superconducting penetration depth
of $\kappa$-(BEDT-TTF)$_{2}$Cu[N(CN)$_{2}$]Br in section~III. Section~IV describes the use of longitudinal 
field measurements to investigate spin fluctuations in the normal state of the material and our conclusions
are presented in section~V. 

\section{Experimental}

Muon-spin spectroscopy\cite{steve} is a sensitive means of probing the superconductivity and local
magnetism of molecular materials of this sort\cite{lee,satoh}. 
Transverse-field (TF) and longitudinal-field (LF)
muon-spin relaxation ($\mu^{+}$SR) measurements
were made on a mosaic sample of $\kappa$-(BEDT-TTF)$_{2}$Cu[N(CN)$_{2}$Br]
at the ISIS facility, Rutherford Appleton Laboratory, UK and the Swiss Muon Source (S$\mu$S),
Paul Scherrer Institut, CH. The polycrystalline sample was made of $\sim 50$
small crystallites which grow as platelets whose large faces are
parallel to the conducting layers (the $ac$ planes for this material).
These were arranged on a piece of Ag foil to cover an area of $\approx 0.5$~cm$^{2}$
so that the conducting layers are parallel to the plane of the mosaic.
TF measurements were made using the MuSR spectrometer at ISIS
where the sample was mounted on a hematite backing plate in a helium cryostat 
with the sample oriented at $45^{\circ}$ to both the applied magnetic 
field and the initial muon spin. In order to reduce any effect of ethylene disorder on 
the superconducting properties, 
the sample was slow cooled at a rate of $\lesssim 5$~K/hour. TF measurements were also 
made using the GPS instrument at S$\mu$S where the sample plane was oriented at 
$90^{\circ}$ to the applied magnetic field and no slow cooling procedure was followed.
LF measurements were made on the ARGUS spectrometer (ISIS) in the latter $90^{\circ}$
geometry. 

\section{TF $\mu^{+}$SR and superconductivity}

TF $\mu^{+}$SR provides a means of
accurately measuring the internal magnetic field distribution in a material,
such as that due to the vortex lattice (VL) in a type II superconductor \cite{sonier}.
In a TF $\mu^{+}$SR experiment spin polarized muons are implanted in the bulk of
a material in the presence of a magnetic field $B_{c1} < B_{\mathrm{a}} < B_{c2}$, 
which is applied perpendicular to the initial muon spin direction. 
Muons stop at random positions on the length scale of the 
VL where they precess about the total local magnetic field $B$ at the muon site 
with frequency $\omega_{\mu} = \gamma_{\mu} B$, where 
$\gamma_{\mu}= 2 \pi \times 135.5$~MHz~T$^{-1}$. 
The observed property of the experiment is the time evolution of the
muon-spin polarization $P_{x}(t)$, which allows the determination of 
the distribution $p(B)$ of local magnetic fields across the sample volume via 
$
P_{x}(t) = \int_{0}^{\infty} p(B) \cos (\gamma_{\mu} B t + \phi) \mathrm{d}B,
$
where the phase $\phi$ results from the detector geometry.

\begin{figure}[htb!]
\begin{center}
\epsfig{file=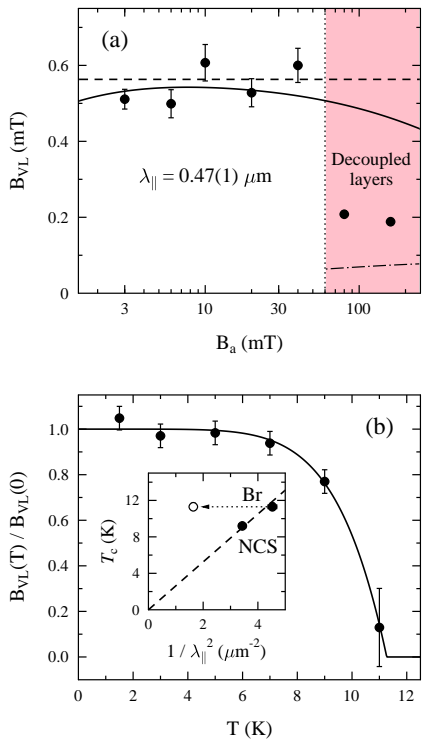,width=5.5cm}
\caption{
(a) Broadening of the 1.5 K TF $\mu^{+}$SR signal due to the VL measured at S$\mu$S. The field $B_{\mathrm{a}}$ was applied perpendicular to the conducting layers.
The dotted line indicates a transition to a decoupled-layer vortex phase (shaded region) occurring between 40 mT and 80 mT. 
The solid line is the fit of the field dependent line width \cite{Brandt03} below the vortex transition, giving $\lambda_{\parallel}$ = 0.47(1) $\mu$m; the dashed line is the corresponding width predicted by Eq.~(\ref{BG}).
The dot-dash line shows the width according to Eq. (\ref{2D}) for fully decoupled vortex layers.
(b) Temperature dependence of the normalised VL broadening measured at ISIS in 3 mT field applied at 45$^{\circ}$ to the sample plane. 
A fit is shown to Eq.~(\ref{one}). 
The inset shows scaling\cite{uemura,blundellpratt} between $T_c$ and $\lambda_{\parallel}^{-2}$, comparing the current data (Br) with that of $\kappa$-(BEDT-TTF)$_{2}$Cu(NCS)$_2$ (NCS)\cite{blundellpratt}; the open circle shifted to the left shows the apparent effect of applying B$_a \ge$~80~mT to Br.  
\label{tffig}}
\end{center}
\end{figure}

Above $T_{c}$ some broadening of the spectrum is caused by randomly directed nuclear 
moments relaxing the muon spins. 
Below $T_{c}$ the spectrum broadens considerably due to the contribution from the VL, but this component was found to be only $\sim 30$\% 
of the total signal for $T<T_{c}$. 
Allowing for part of the non-superconducting signal to come from the cryostat and sample mounting material, 
we estimate that at most 50\% of the sample is giving a superconducting signal.
This is significant since it suggests 
that the majority of muons are not stopped in regions where there is a well defined
VL and may indicate
the coexistence of a competing phase in this material. 
In fact,  there is evidence for
a phase separation between superconducting and insulating regions in this material\cite{taylor}
which is strongly dependent on disorder, such as that introduced through rapid cooling.
The non-superconducting contribution to our signal for $T<T_{c}$ decreases weakly and 
linearly with increasing temperature.
Rapid cooling of the sample was found to reduce the superconducting volume fraction still further.
With such a small superconducting fraction it is difficult to accurately extract detailed lineshape information
from our data, however the RMS width of the VL field distribution $B_{\mathrm{VL}}$ 
can be straightforwardly obtained by fitting the data to the sum of two Gaussian relaxation components, reflecting the superconducting VL and non-superconducting contributions. These are easily separated below $T_c$ since the contribution from the VL has a much larger relaxation rate.
The field dependence of the VL component is shown in Fig.~\ref{tffig}(a).
The low field data (40 mT and below) fit well to the width expected from the solution of a Ginzburg-Landau (GL) model of the triangular VL field distribution \cite{Brandt03} with an in-plane penetration depth of 0.47(1) $\mu$m. $B_{\mathrm{VL}}$ is not very sensitive to $B_{c2}$ in this field range, so we assume the reported value \cite{Wosnitza} $B_{c2}$=10(2)~T; the GL parameter $\kappa$ is then estimated to be 80(10). 
The penetration depth in the highly conducting planes can also be estimated via the approximate expression for a high anisotropy superconductor 
\begin{equation}
\lambda_{\parallel} \approx \left(0.00371 \right)^{\frac{1}{4}}
\left[\frac{\Phi_{0} \cos\theta}{B_{\mathrm{VL}}} \right]^{\frac{1}{2}},
\label{BG}
\end{equation}
where $\Phi_{0}$ is the flux quantum and $\theta$ is the angle of the applied field with respect to the normal to the plane. 
This field-independent expression is valid at intermediate fields $B_{c1}{\ll}B_a{\ll}B_{c2}$ and the $B_{\mathrm{VL}}$ value obtained from 
Eq.~(\ref{BG}) for $\lambda_{\parallel}$ = 0.47 $\mu$m is shown as the dashed line in Fig.~\ref{tffig}(a); this is seen to be reasonably close to the field-dependent GL width in a field region centred around 10~mT.  

\begin{table*}[htb]
\caption{Comparison of penetration depth $\lambda_{\parallel}$ extracted from $\mu^{+}$SR measurements
under different experimental conditions.}
\label{table1}
\begin{ruledtabular}
\begin{tabular}{c|ccccccc|}
Anion &Study &Skewness&$B_{\mathrm{a}}$~(mT) &  $\lambda_{\parallel}$~($\mu$m)  \\ 
\hline
Cu[N(CN)$_{2}$Br & This study& 0 & 3-40        &  0.47(1)  \\
Cu[N(CN)$_{2}$Br & This study&  0.7& 3      & 0.47(1)  \\
Cu[N(CN)$_{2}$Br & This study& 0 & 160        & 0.82(1)   \\
Cu[N(CN)$_{2}$Br & Ref.~\onlinecite{le}& 0& 300   & 0.78    \\
Cu(SCN)$_{2}$ &  Ref.~\onlinecite{lee}&0.38 &  2.5  &  0.54(2)   \\
Cu(SCN)$_{2}$ &  Ref.~\onlinecite{harshman} &0 & 13-400       &    0.77(7)   \\
\end{tabular}
\end{ruledtabular}
\end{table*}

Our value of $\lambda_{\parallel}$ is rather smaller than previous reports, e.g.\ the range 
$0.57 \leq \lambda_{\parallel} \leq 0.69~\mu$m obtained from reversible high field magnetization measurements\cite{yoneyama},
and it is significantly lower than the value $\lambda_{\parallel} \approx 0.78~\mu$m estimated from a previous
$\mu^{+}$SR measurement \cite{le} in a field of 300~mT. 
These differences can be understood by noting that the measured 
linewidth decreases sharply for $B_{\mathrm{a}}\geq 80$~mT (Fig.~\ref{tffig}(a)). 
If this suppressed linewidth were taken to represent a full 3D VL, a penetration depth of order
$\lambda=0.80~\mu$m would also be obtained from our data using Eq. (\ref{BG}), in good agreement with the previous $\mu^{+}$SR result.
The decrease in $B_{\mathrm{VL}}$ with applied field that we observe
is not surprising, given the vortex phase diagram in (BEDT-TTF)$_{2}$Cu(SCN)$_{2}$\cite{lee,pratt}. 
In that material the ideal 3D triangular VL which exists at low temperatures and low fields \cite{Vinnikov} is destroyed by the
application of fields above a disorder-dependent threshold, that can be as low as 6~mT, causing a transition to a decoupled-layer vortex glass phase, 
accompanied by a sharp decrease in the measured linewidth \cite{pratt}. It is therefore 
likely that a similar transition to a decoupled-layer vortex glass phase occurs in the $X=$Cu[N(CN)$_{2}$Br material
above a theshold field in the 40--80~mT region.
The effect on the width of the field distribution in the limit of losing all 
inter-layer correlation has been calculated\cite{Harshman93,Brandt95}, giving 
\begin{equation}
B_{\mathrm{2D}}/B_{\mathrm{VL}}=1.4(s/a)^{1/2},
\label{2D}
\end{equation}
where $s$ is the inter-layer spacing and $a$ is vortex spacing within the layers. 
This limit is shown as the dot-dash line in Fig.~\ref{tffig}(a) and it can be concluded that some residual interlayer correlations remain here since the data points lie significantly above this limit. 
A further feature of our data in the decoupled vortex phase is that the superconducting fraction recovers to the larger value obtained in measuremants made under slow-cooling conditions, suggesting that the disorder-induced phase separation is becoming significantly reduced once the layers are decoupled.

The temperature dependence of $B_{\mathrm{VL}}$ has been measured in a field of 3 mT applied in a 45$^{\circ}$ geometry and this is shown in Fig.~\ref{tffig}(b).
This can be fitted to an empirical power law 
\begin{equation}
B_{\mathrm{VL}}(T) = B_{\mathrm{VL}}(0) [1-(T/T_{c})^{r}],
\label{one}
\end{equation}
where $B_{\mathrm{VL}}(0)$ is the zero temperature contribution from the VL.
The VL here therefore appears to thermally stable, in contrast to the case of (BEDT-TTF)$_{2}$Cu(SCN)$_{2}$ where a clear melting of the VL is seen \cite{pratt05}.
The values $B_{\mathrm{VL}}(0)$=0.31(1)~mT and $T_c$=11.3(4)~K are obtained from the fit.
The penetration depth in the highly conducting planes could be obtained by 
taking $\theta = 45^{\circ}$ in Eq.~(\ref{BG}), 
however the absolute accuracy is reduced compared to 90$^{\circ}$ measurements for two reasons: 
firstly the width measured at $45^{\circ}$ is particularly sensitive to any angular misalignment between the sample and field and secondly the angular scaling in highly anisotropic superconductors can be much more complex than the simple form implied by Eq.~(\ref{BG}) \cite{pratt2}.
Returning to the 90$^{\circ}$ data, we note that a marginally better fit is obtained by allowing for an asymmetric lineshape \cite{lee} 
parameterized using the skewness parameter $\beta = (\langle B \rangle - B_{\mathrm{a}})/B_{\mathrm{VL}}$.
The optimum fit is achieved with $\beta = 0.7$, which is reasonably close to the ideal triangular lattice value of 0.60, that was observed in more detailed studies of the vortex phases of the X=Cu(NCS)$_{2}$ compound \cite{pratt}. 
This fit also yields 0.47(1)~$\mu$m for $\lambda_{\parallel}$.
The values obtained here and in previous muon studies for 90$^{\circ}$ geometry are summarized in Table~\ref{table1}. 
We may compare our values of $\lambda_{\parallel}$ and $T_{c}$
for the X=Cu[N(CN)$_{2}$]Br material to values
obtained previously for the X=Cu(SCN)$_{2}$ material. This is shown as
the inset to Fig.~\ref{tffig}(b) and suggests that there is a scaling relation between these two parameters.
Although this pair of $\kappa$-phase points lie rather close, a simple linear scaling \cite{uemura} would be reasonably consistent with the data. 
More generally, a $T_c \propto \lambda^{-3}$ scaling has been suggested to apply when a broader range of molecular superconductors is considered \cite{blundellpratt}, but more data would be required to test whether this holds or not within the subset of $\kappa$-phase BEDT-TTF compounds.

\section{LF $\mu^{+}$SR and spin fluctuations}

We now turn to the local magnetic fluctuations in the normal state of this system,
which we probe with
LF $\mu^{+}$SR measurements made in a field of $B=2$~mT. In these measurements
the initial muon spin is directed parallel to the applied field and we measure the 
polarization $P_{z}(t)$ along the same direction via the muon asymmetry function $A(t)[\propto P_{z}(t)]$. 
Dynamics in the local magnetic field distribution will cause muon spins to flip
leading to a depolarization with relaxation rate $\Lambda$. 
The small applied longitudinal field was intended to quench the contribution to the spectra 
of background from nuclear moments allowing the contribution due to electron spin dynamics to be 
discerned. 

\begin{figure}[htb]
\begin{center}
\epsfig{file=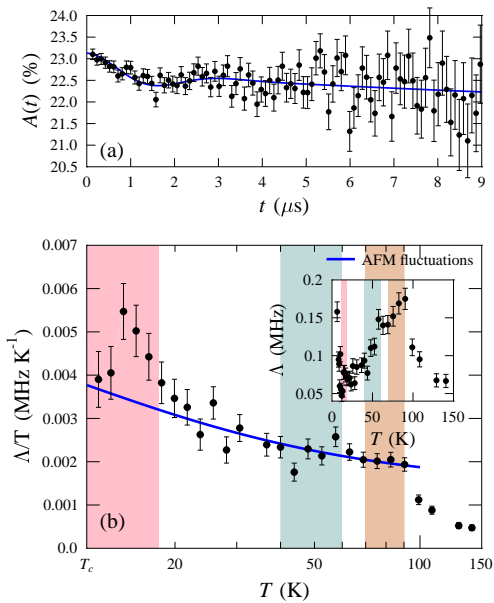,width=7cm}
\caption{(a) Typical spectra measured in a LF of $B=2$~mT at $T=20$~K.
(b) Temperature evolution of $\Lambda/T$. Regions~I, II and III
are shaded and a fit to a model of antiferromagnetic spin fluctuations is shown. 
{\it Inset:}
Evolution  of $\Lambda$ with $T$ extracted from fits to Eq.~(\ref{fitlf}).
\label{lambdat}}
\end{center}
\end{figure}

A typical LF $\mu^{+}$SR spectrum measured at 20~K is shown in Fig.~\ref{lambdat}(a). 
The relaxing amplitude is quite small, but this is not unexpected given previous zero field
muon results measured on a related BEDT-TTF material \cite{satoh}. In that study, muons were shown to 
be sensitive to magnetic order within the organic layers but only a small relaxing asymmetry was observed.
We therefore expect that our results will reflect magnetic fluctuations within the organic layers of 
$\kappa$-(BEDT-TTF)$_{2}X$.
Above 100~K our spectra are best described with a single exponential function 
with relaxation rate $\Lambda$
characteristic of dynamic fluctuations in the local magnetic field at the muon sites in the material. 
For $T<100$~K, there is a single, highly damped, temperature
dependent oscillation visible for times $t \leq 4~\mu$s.
In this context, spectra of this form suggest a contribution from quasistatic 
disordered magnetic moments
which is typically described by a Kubo-Toyabe (KT) function\cite{hayano} $f_{\mathrm{KT}}(\Delta,B,t)$ 
where the data are best fitted with a constant field width $\Delta=0.72$~MHz for all $T<100$~K.
We note that this is quite large for randomized nuclear fields (where typically $\Delta < 0.3$~MHz)
and given the small size of the electronic moments observed with muons in BEDT-TTF layers  previously
\cite{satoh} it may be that this signal results from a population of static disordered electronic moments. 
The origin of such quasistatic electronic moments is unclear although 
we note that there is no decrease in $\Delta$ with increasing $T$ as would occur in the presence of 
magnetic order. We note that it is possible that this signal arises from those portions of the sample
not contributing to the superconducting VL in TF measurements. 

The dynamic contribution to the spectra for $T< 100$~K was found to be best 
modelled by multiplying the KT function by an exponential factor $e^{-\Lambda t}$ and
the data were fitted to the resulting functional form
\begin{equation}
A(t) = A_{0} f_{\mathrm{KT}}(\Delta,B,t) e^{-\Lambda t} + A_{\mathrm{bg}}.
\label{fitlf}
\end{equation}
The extracted relaxation rates $\Lambda$ are shown in Fig.~\ref{master}(a) and inset in Fig.~\ref{lambdat}(b).
Other parameterizations are possible but lead to less successful fits. However, for comparison, a fit across 
the measured temperature regime of a single
exponential with relaxation rate $\Lambda$ is shown in Fig.~\ref{master}(a), where we see that 
both track the same behavior.

At temperatures $T<100$~K, the dominant trend  is a roughly 
linear increase in relaxation rate with $T$ up to
around 100~K, above which it drops precipitously [inset Fig.~\ref{lambdat}(b)]. 
The linear regime observed for $T<100$~K is typical of
a Korringa law relaxation \cite{blundellcox}, resulting from flip-flop transitions
of electronic and muon spins, and which is expected to 
lead to a relaxation rate
$\Lambda = \frac{1}{T_{1}} \propto k_{\mathrm{B}} T \sum_{q} A(\boldsymbol{q})^{2} \lim_{\omega \rightarrow 0} 
\chi''(\boldsymbol{q},\omega)/\omega$,
where $\chi''(\boldsymbol{q},\omega)$ is the imaginary part of the dynamic magnetic
susceptibility and $A(\boldsymbol{q})$ is the hyperfine coupling. 
For the case of a $\boldsymbol{q}$-independent coupling, 
we obtain $\lambda \propto k_{\mathrm{B}} T [A \chi(0,0)]^{2}$, showing that the
muon relaxation rate probes the local spin susceptibility $\chi(\boldsymbol{q}=0,\omega=0)$.
The sharp decrease in relaxation rate around 100~K broadly coincides with region I, which
may suggest a link with the ethylene disorder effects 
and glass-like transition in this region. 
Given the Korringa law behavior observed at low temperatures,
it is likely that the crossover to nonmetallic behavior of the system  near region I 
[Fig.~\ref{master}(b)] changes the nature of
the muons' coupling to the material and that at these elevated temperatures 
a different, dominant relaxation channel presumably opens up. 
To further investigate the temperature evolution 
of the local susceptibility from the Korringa relaxation we plot $\Lambda/ T (\equiv 1/T_{1}T)$ against $T$ 
in Fig.~\ref{lambdat}(b). We see that on warming above $T_{c}$ the quantity $\Lambda/T$ 
peaks at around 15~K before decreasing sharply around region~III
($15 < T \lesssim 25$~K) indicating a significant decrease of the 
local magnetic susceptibility in this region. Above 30~K $\Lambda/T$ decreases only gradually
throughout region II until dropping off suddenly in region I.

Our muon $\Lambda/T$ results are quite different to the NMR $1/T_{1}T$ results [Fig.\ref{master}]
where a parger peak is seen in region II. 
We do not see any dramatic change of behavior in 
region II in the $\mu^{+}$SR results, while 
there is no suggestion of any anomaly in the NMR behavior in regions~I and III. 
In fact, the NMR results for $50$~K$< T <300$~K are well described by a model
of antiferromagnetic spin fluctuations\cite{yusuf} based on the  MMP model\cite{mmp}, which
predicts
$1/T_{1}T \approx A + 
B/(T/T_{x}+1),$
where $A$ and $B$ are constants that depend on the correlation length of the
spin fluctuations whose energy scale is determined by the temperature parameter $T_{x}$.
As shown Fig.~\ref{lambdat}(b) the model fits the muon $\Lambda$ only passably 
well for $T<100$~K becomes gradually worse as region~III  
approached from above.
However, given that the
muon sites in this material are likely to be different to the $^{13}$C site where the nuclear resonance
is achieved\cite{miyagawa}, it is possible that the muon will probe a quite different field distribution
to that probed in NMR.
We note further that our muon results are also quite different
to the LF $\mu^{+}$SR behavior observed in molecular superconductors such as the 
alkali-fullerides \cite{macfarlane} where $\Lambda/T$ is quite featureless above $T_{c}$. 

\section{Conclusions}

In conclusion, our investigation of the superconducting and normal state properties
of  $\kappa$-(BEDT-TTF)$_{2}$Cu[N(CN)$_{2}$]Br using implanted muons has allowed us to
identify a number of experimental features of this system.
In the superconducting phase two important effects are seen: firstly, a reduced superconducting signal fraction that is affected by sample cooling rate and is consistent with the phase separation suggested by earlier studies and secondly, a vortex transition taking place between 
40~mT and 80~mT, that reduces the width of the magnetic field distribution for higher measurement fields.
By taking these two effects carefully into account we have obtained an improved and more reliable estimate 
for the $T=0$ in-plane penetration depth $\lambda_{\parallel}$=0.47(1) $\mu$m and find that the trend of
 $T_c$ increasing with 
superfluid stiffness $\rho_s \propto 1/\lambda_{\parallel}^2$ for $\kappa$-(BEDT-TTF)$_2$X superconductors is consistent with the overall trend for molecular superconductors, whereas previous data had suggested that $\kappa$-(BEDT-TTF)$_{2}$Cu[N(CN)$_{2}$]Br was anomalous in this regard. 
In the normal state we find a large peak in the
longitudinal muon spin relaxation rate around 100~K, coinciding with the region where 
ethylene disorder effects are prevalent. 
We also observe an enhancement of the local magnetic spin susceptibility above $T_{c}$ where vortex fluctuations
occur in Nernst measurements. 
It would be desirable in future to extend $\mu^{+}$SR studies of the normal state magnetic properties
to the other members of this series in order to follow how these features evolve across the phase diagram. 

We are grateful to Alex Amato, Andrew Steele and Peter Baker for
experimental assistance and to EPSRC (UK) for financial support.
Part of this work was performed at S$\mu$S and part at the STFC ISIS facility and we
are grateful to PSI and STFC for the provision of beamtime.
Work supported by U.\ Chicago Argonne, LLC, Operator of Argonne National Laboratory
(``Argonne"). Argonne, a U.S. Department of Energy Office of Science
laboratory, is operated under Contract No.\ DE-AC02-06CH11357.


\begin{thebibliography}{xx}

\bibitem{ishiguro}
T. Ishiguro, K. Yamaji and G. Saito {\it Organic Superconductors: 2nd edition} (Springer, Berlin)
2006.

\bibitem{powell}
B.J. Powell and R.H. McKenzie, J. Phys.: Condens. Matter {\bf 18}, R827 (2006).

\bibitem{sano}
K. Sano, T. Sasaki, N. Yoneyama and N. Kobayashi,
Phys. Rev. Lett. 104, 217003 (2010).

\bibitem{taylor}
O.J. Taylor, A. Carrington and J.A. Schlueter, 
Phys. Rev. B {\bf 77}, 060503(R) (2008).

\bibitem{yusuf}
E. Yusuf, B.J. Powell, R.H. McKenzie, Phys. Rev. B {\bf 75}, 214515 (2007).

\bibitem{nam}
M.-S. Nam, A. Ardavan, S.J. Blundell, J.A. Schlueter, Nature {\bf 449}, 584 (2007).

\bibitem{wolter}
A.U.B. Wolter, R. Feyerherm, E. Dudzik, S. S\"{u}llow, C. Strack, M. Lang and D. Schweitzer,
Phys. Rev. B {\bf 75}, 104512 (2007).

\bibitem{kagawa}
F. Kagawa, M. Miyagawa and K. Kanoda, Nature {\bf 436}, 534 (2005). 

\bibitem{williams}
J.M. Williams {\it et al.}, Inorg. Chem. {\bf 29}, 3272 (1990). 

\bibitem{muller}
J. M\"{u}ller, M. Lang, F. Steglich, J.A. Schlueter, A.M. Kini and T. Sasaki, Phys. Rev. B {\bf 65}, 
144521 (2002). 

\bibitem{yu}
R.C. Yu, J.M. Williams, H.H. Wang, J.E. Thompson, A. M. Kini, K. D. Carlson, J. Ren, M.-H. Whangbo
and P.M. Chaikin, Phys. Rev. B {\bf 44}, 6932 (1991).

\bibitem{miyagawa}
K. Miyagawa, K. Kanoda and A. Kawamoto, Chem. Review, {\bf 104}, 5635 (2004). 

\bibitem{steve}
S.J. Blundell, Contemp. Phys. {\bf 40}, 175 (1999).

\bibitem{lee}
S.L. Lee, F.L. Pratt, S.J. Blundell, C.M. Aegerter, P.A. Pattenden,
K.H. Chow, E.M. Forgan, T. Sasaki, W. Hayes and H. Keller,
Phys. Rev. Lett. {\bf 79}, 1563 (1997).

\bibitem{satoh}
K. Satoh, H. Taniguchi, A. Kawamoto and W. Higemoto,
Physica B {\bf 374-375}, 99 (2006).

\bibitem{sonier}
J.E. Sonier, J.H. Brewer, and R.F. Kiefl, Rev. Mod. Phys. {\bf 72}, 769 (2000).

\bibitem{Brandt03}
E.~H. Brandt, {Phys. Rev. B} {\bf 68}, {054506} (2003).

\bibitem{Wosnitza}
H. Elsinger, J. Wosnitza, S. Wanka, J. Hagel, D. Schweitzer, and W. Strunz, Phys. Rev. Lett. {\bf 84}, 6098 (2000). 

\bibitem{yoneyama}
N. Yoneyama, A. Higashihara, T. Sasaki, T. Nojima and N. Kobayashi,
J. Phys. Soc. Jpn. {\bf 73}, 1290 (2004). 

\bibitem{le}
L.P. Le, G.M. Luke, B.J. Sternlieb, W.D. Wu, Y.J. Uemura, 
J.H. Brewer, T.M. Riseman, C.E. Stronach, G. Saito, H. Yamochi,
H.H. Wang, A.M. Kini, K.D. Carlson, J.M. Williams, Phys. Rev. Lett
{\bf 68}, 1923 (1992).

\bibitem{pratt}
F.L. Pratt, S.L. Lee, C.M. Aegerter, C. Ager, S.H. Lloyd, S.J. Blundell,
F.Y. Ogrin, E.M. Forgan, H. Keller, W. Hayes, T. Sasaki, N. Toyota and S. Endo,
Synth. Met. {\bf 120}, 1015 (2001).

\bibitem{Vinnikov}
L.Ya.~Vinnikov, T.L.~Barkov, M.V.~Kartsovnik and N.D.~Kushch,
Phys. Rev. {\bf B} 61, 14358 (2000).


\bibitem{Harshman93}
D.R. Harshman, E.H. Brandt, A.T. Fiory, M. Inui, D.B. Mitzi, L.F. Schneemeyer and J.V. Waszczak
Phys. Rev. {\bf B} 47, 2905 (1993). 

\bibitem{Brandt95}
E.H. Brandt, Rep. Prog. Phys. {\bf 58}, 1465 (1995).


\bibitem{pratt05}
F.L. Pratt, S.J. Blundell, T. Lancaster, M.L. Brooks, S.L. Lee, N. Toyota and T. Sasaki,
Synth. Met. {\bf 152}, 417 (2005).

\bibitem{pratt2}
F.L. Pratt, I.M. Marshall, S.J. Blundell, A. Drew, S.L. Lee, F.Y. Ogrin,
N. Toyota and I. Watanabe, Physica B, {\bf 326}, 374 (2003).

\bibitem{uemura}
Y.~J. Uemura {\em et al},
{Phys.\ Rev.\ Lett.}  {\bf 62}, {2317}
  (1989);
Y.~J. Uemura {\em et al},
 {Phys.\ Rev.\ Lett.}  {\bf 66}, {2665} (1991).

\bibitem{blundellpratt}
F.L. Pratt and S.J. Blundell, Phys. Rev. Lett {\bf 94}, 097006 (2005). 

\bibitem{harshman}
D.R. Harshman, A.T. Fiory, R.C. Haddon, M.L. Kaplan, T. Pfiz,
E. Koster, I. Shinkoda and D.Ll. Williams, 
Phys. Rev. B {\bf 49}, 12990 (1994).
\bibitem{hayano}
R.S. Hayano, Y. J. Uemura, J. Imazato, N. Nishida, T. Yamazaki and R. Kubo,
 Phys. Rev. B {\bf 20}, 850 (1979).

\bibitem{blundellcox}
S.J. Blundell and S.F.J. Cox, J. Phys.: Condens. Matter {\bf 13}, 2163 (2001).

\bibitem{mmp}
A.J. Millis, H. Monien and D. Pines, Phys. Rev. B {\bf 42}, 167 (1990). 

\bibitem{macfarlane}
W.A. MacFarlane, R.F. Kiefl, S. Dunsiger, J.E. Sonier, J. Chakhalian, J.E. Fischer, T. Yildirim
and K.H. Chow, Phys. Rev. B {\bf 58}, 1004 (1998).

\end{thebibliography}
\end{document}